\documentclass[12pt]{sn-jnl}

\usepackage{graphicx} 
\usepackage{amsmath}
\usepackage{amssymb}
\usepackage{tikz}
\usetikzlibrary{shapes,arrows}
\usepackage{parskip}
\usepackage{comment}
\usepackage{url}
\usepackage{hyperref}

\usepackage{subcaption}

\usepackage{graphicx}
\usepackage{overpic}

\newtheorem{theorem}{Theorem}
\usetikzlibrary{positioning, arrows.meta}

\tikzstyle{block} = [draw, fill=white, rectangle, minimum height=3em, minimum width=6em]
\tikzstyle{sum} = [draw, fill=white, circle, node distance=1cm]
\tikzstyle{input} = [coordinate]
\tikzstyle{output} = [coordinate]
\tikzstyle{pinstyle} = [pin edge={to-,thin,black}]

\begin{document}

\title[Co-spreading dynamics of smoking behavior and awareness]{Co-spreading dynamics of smoking behavior and awareness on social contact networks}

\author{\fnm{Saicharan Ritwik} \sur{Chinni}}

\author*{\fnm{Anupama} \sur{Sharma}\thanks{Corresponding author}}
\email{anupamas@goa.bits-pilani.ac.in}

\affil{\orgdiv{Department of Mathematics},
\orgname{Birla Institute of Technology and Science, Pilani, K K Birla Goa Campus, Zuarinagar},
\orgaddress{\city{Sancoale}, \postcode{403726}, \state{Goa}, \country{India}}}

\abstract{Smoking behavior and awareness co-spread through social interactions, giving rise to coupled contagion processes on social contact networks. In addition to initiation and cessation, awareness of the harmful effects of smoking plays an important role in shaping individual behavior and population-level outcomes. In this work, we develop a mathematical model to study the coupled dynamics of smoking behavior, quitting, and awareness in a population. A deterministic framework based on ordinary differential equations is first formulated to capture the interplay between social influence and awareness-driven behavioral change. Analysis of the model reveals the existence of smoking-free and smoking-endemic steady states, and identifies conditions under which awareness can reduce or suppress the persistence of smoking. Since social interactions are often localized rather than well mixed, the mean-field description is then extended to a network-based model that incorporates structured contact patterns. Numerical simulations performed on empirical social networks indicate that contact heterogeneity and localized awareness spreading can influence the effectiveness of interventions. Our findings underscore the importance of population structure when devising awareness-based intervention strategies for smoking cessation. }


\maketitle

\section{Introduction}
Smoking remains one of the leading preventable causes of death worldwide~\cite{WHO2025,GBD2019Tobacco2021}, motivating extensive research into its dynamics and control. Empirical research shows that smoking behavior among peers is significantly associated with an individual’s own smoking uptake and continuation. For example, adolescents with smoking friends are about twice as likely to begin and continue smoking compared to those without such ties, and social network characteristics such as friendship ties and peers’ smoking behavior reliably predict smoking outcomes over time~\cite{Liu2017}. Not only smoking behavior but also awareness of its health risks can spread through social interactions, and the structure of social networks can shape the effectiveness of awareness-based interventions~\cite{Jia2024}. These findings suggest that smoking behavior and awareness co-spread through interpersonal contact, forming coupled contagion-like processes. Mathematical modeling has therefore been widely used to investigate how individual-level contact structure gives rise to collective patterns of smoking prevalence and behavioral change.  

Early modeling efforts drew inspiration from epidemic theory, representing smoking initiation and cessation as contagious processes  similar to the disease transmission~\cite{Rowe1992,Castillo1997,Sharomi2008}. These models typically assume homogeneous mixing and describe transitions between nonsmokers, smokers, and former smokers using systems of ordinary differential equations (ODEs). Recent ODE models have incorporated additional compartments and nonlinear transitions, and have further introduced fractional and time-fractional nonlinear smoking growth rates to better capture the complexity of smoking behavior~\cite{Guerrero2013,Pang2015,Rahman2018, Khan2019,Zhang2019,Ahmad2022,Sanchez2023,Pavani2024}. For example, age-structured model of smoking dynamics have been proposed to account for demographic heterogeneity and to investigate optimal control strategies under heterogeneous risk profiles~\cite{Rahman2018}. More recently, Zhang et al. (2024) developed a five-dimensional smoking epidemic model that accounts for different degrees of smoking (such as casual and heavy smoking) and incorporates relapse after quitting \cite{Zhang2024}. This model extended the traditional SIR-type approach by introducing separate compartments for varying smoker categories and analyzing both deterministic dynamics and stochastic perturbations. While such mean-field formulations provide analytical insight, they neglect the fact that social interactions are structured rather than random and that behavioral change is mediated by localized peer influence.

Heterogeneous modeling approaches adopt a bottom-up approach to simulating smoking behavior by modeling individual agents (people) with rules for behavior change and interactions. Such frameworks can capture how interventions affect different subgroups and how smoking behavior spreads unevenly across a population. These ideas have been applied, for example, to study the concurrent dynamics of smoking and e-cigarette use and their mutual influence. Qin et al. (2019) developed an individual-level simulation model to examine how e-cigarette uptake may alter smoking initiation and cessation patterns~\cite{Qin2019}. The role of peer influence in smoking uptake and cessation is well documented in empirical studies. Classic work by Christakis and Fowler (2008) analyzed a large social network and found that if a person quits smoking, the chances that their friends quit increase significantly, leading to clusters of quitting spreading through the network \cite{Christakis2008}. Similarly, smoking initiation often occurs in friendship groups, where social norms and peer pressure play a decisive role. Adolescents with a higher proportion of smoking friends are far more likely to start smoking themselves, and vice versa for those with non-smoking friends \cite{Kobus2003,SimonsMorton2010}. These observations imply that the structure of social interactions can critically shape smoking prevalence over time: dense clusters of smokers can reinforce and sustain smoking behavior, while shifts toward non-smoking within a cluster can generate self-reinforcing norms in the opposite direction.

Guided by these observations, we formulate a model of coupled behavioral and awareness dynamics driven by peer interaction. Individuals transition between behavioral states in a manner analogous to infection and recovery processes, with transition rates shaped by peer effect. Prior work on coupled contagion processes has shown that awareness can spread concurrently with a primary behavior or disease and influence its dynamics~\cite{Funk2009,Granell2013}. However, existing smoking models that incorporate awareness typically represent it as a population-level or externally driven influence, rather than as a quantity that propagates through interpersonal contact. In particular, although the effects of awareness on smoking initiation and cessation have been explored in previous compartmental frameworks~\cite{Sharma2015, Sharma2020, Sofia2023}, the mechanism by which awareness spreads - namely through localized, peer-to-peer interactions - has received much less attention. In real-social settings, individuals who quit smoking may actively influence their contacts, promoting awareness and encouraging further cessation through word-of-mouth communication. This socially mediated mechanism is supported by randomized trials showing that smoking cessation can be reinforced through peer-to-peer interaction in online social networks \cite{Pechmann2017}. This localized mode of transmission implies that social contact structure can play a decisive role in shaping both smoking and awareness dynamics. This localized transmission mechanism suggests that awareness dynamics are intrinsically linked to social contact structure. Since network-based epidemic models have demonstrated that contact heterogeneity can substantially alter spreading dynamics \cite{PastorSatorras2001,Newman2003,KeelingEames2005}, accounting explicitly for how awareness propagates through social interactions becomes essential for understanding smoking dynamics driven by peer influence.

To address these gaps, we develop a modeling framework that integrates awareness as an explicit dynamical compartment interacting with smoking uptake, cessation, and relapse. We first analyze a mean-field formulation and derive results, including a behavior-reproduction number and local stability conditions for equilibria, and investigate parameter sensitivities to identify key intervention leverage points. This mean-field formulation provides analytical tractability and allows us to derive key dynamical properties of the system. We then extend the mean-field model to a network-based formulation using degree-block approximations in order to capture the impact of contact heterogeneity on long-term outcomes. Through simulations on both theoretical and empirical networks, we demonstrate how structural features of social contact patterns influence the prevalence of smoking and awareness. This combined analytical and network-based perspective links individual-level social influence, awareness propagation, and population-level behavioral change within a single coherent modeling framework. The rest of the paper is organized as follows: Section~2 describes the formulation of awareness-coupled compartmental model for smoking; Section~3 presents its theoretical analysis, including equilibria and their stability; Section~4 contains numerical simulations of the model; Section~5 develops and analyzes the network-based modelling framework; and Section~6 discusses implications and future directions.

\section{The Model} 

We formulate a deterministic compartmental model to describe the coupled dynamics of smoking behavior, cessation, relapse, and awareness in a population. Individuals are classified into four mutually exclusive compartments: non-smokers ($N$), smokers ($S$), quitters ($Q$), and aware individuals ($A$). Transitions between compartments are governed by peer influence, awareness-mediated behavior change, relapse, and demographic turnover. Smoking initiation is modeled as a contact-driven process between smokers and non-smokers, while cessation is facilitated through interactions between smokers and aware individuals. Quitters may relapse into smoking due to social exposure, and awareness increases both through quitting and external recruitment, such as public health interventions. The resulting system of nonlinear ordinary differential equations provides a mean-field description of these processes and serves as the baseline framework for subsequent analytical and network-based extensions. The schematic representation of the model is shown in Figure~1, and the governing equations are derived below.

Consider a region with total population $T$ at any time $t.$ The whole population is divided into four classes: individuals who do not smoke cigarettes, Non-smokers ($N$); individuals who are regular smokers with non-zero FTND-score, Smokers ($S$), individuals who has quit smoking, Quitter ($Q$) and the individuals who are aware of ill-effect of smoking and aware others too, Aware ($A$).


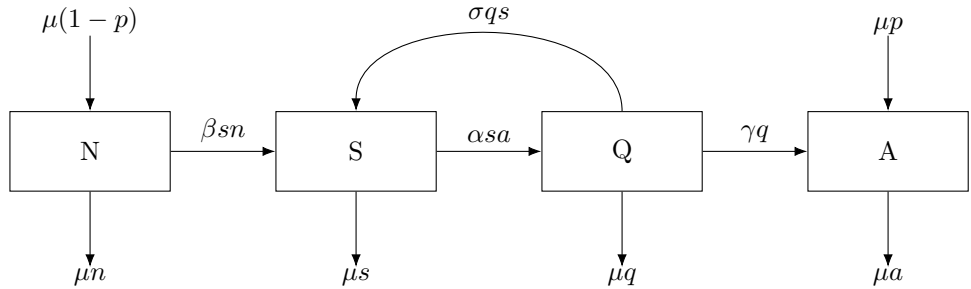
\begin{figure}[htpb]
\centering
\begin{tikzpicture}[auto, node distance=3cm,>=Latex]
    \node [input, name=input] {};
    \node [block, right of=input] (N) {N};
    \node [block, right of=N, node distance=3.5cm] (S) {S};
    \node [block, right of=S, node distance=3.5cm] (Q) {Q};
    \node [block, right of=Q, node distance=3.5cm] (A) {A};

    \draw [->] (N.south) -- node[below = 0.4cm] {$\mu n$} ++ (0,-1cm);
    \draw [->] (S.south) -- node[below = 0.4cm] {$\mu s$} ++ (0,-1cm);
    \draw [->] (Q.south) -- node[below = 0.4cm] {$\mu q$} ++ (0,-1cm);
    \draw [->] (A.south) -- node[below = 0.4cm] {$\mu a$} ++ (0,-1cm);
    \draw [->] (Q) to[out=90, in=90, looseness=1] node[above] {$\sigma qs$} (S);
    \draw [->] ([yshift=1cm]A.north) -- node[above = 0.4cm] {$\mu p$} ++(0,-1cm);
    \draw [->] ([yshift=1cm]N.north) -- node[above = 0.4cm] {$\mu (1-p)$} ++(0,-1cm);
    \draw [->] (N) -- node[name=u1] {$\beta sn$} (S);
    \draw [->] (S) -- node[name=u2] {$\alpha sa$} (Q);
    \draw [->] (Q) -- node[name=u3] {$\gamma q$} (A);
\end{tikzpicture} 
\caption{Schematic diagram of the model}
\end{figure}

The following is the system of nonlinear ordinary differential equations:\\

\begin{equation}
\begin{aligned}
\frac{dN}{dt} &= \mu(1-p)T - \frac{\beta SN}{T} - \mu N \\
\frac{dS}{dt} &= \frac{\beta SN}{T} - \frac{\alpha SA}{T} + \frac{\sigma QS}{T} - \mu S\\
\frac{dQ}{dt} &= \frac{\alpha SA}{T} - \frac{\sigma QS}{T} - \gamma Q - \mu Q \\
\frac{dA}{dt} &= \mu pT + \gamma Q - \mu A
\end{aligned}
\end{equation}
where, N(0) = $\mathrm{N}_0$, S(0) = $\mathrm{S}_0$, Q(0) = $\mathrm{Q}_0$, A(0) = $\mathrm{A}_0$. 

In the above system, $\mu$ denotes both the per capita influx rate of individuals of smoking age and the mortality rate, which are assumed to be identical. The constant $\beta$ represents the effective contact rate, defined as the product of the average number of influential contacts per unit time for a non-smoker and the probability of adopting smoking behavior upon contact with a smoker. Aware individuals disseminate awareness via word-of-mouth among smokers at a rate $\alpha$. The constant $\sigma$ represents the per capita contact-driven relapse rate due to peer influence, while $\gamma$ denotes the per capita rate at which quitters become aware, in the sense that they begin educating others about the harmful effects of smoking. All parameters are assumed to be positive. The parameter $p$ represents the fraction of individuals who enter the population already aware and willing to promote awareness, with $0<p<1$.

Combining the four equations of model system (1) gives dT/dt = 0, which implies that the total population T is constant. Since system~(1) describes human population dynamics, all state variables and parameters are assumed to be nonnegative for all $t \ge 0$. We therefore analyze system~(1) on the positively invariant set
\[
\Omega = \left\{ (N, S, Q, A) \in \mathbb{R}_+^4 \;:\; 0 \le N, S, Q, A \le T \right\},
\]
which constitutes the region of attraction of the system.

\section{Model Analysis}

\subsection{Steady states}

As the total population, $N+S+Q+A = T$ is a constant over time, 
it is useful to consider the fractions , $N/T = n, S/T = s, Q/T = q$ and $A/T = a,$
with $n+s+q+a = 1.$ So the above system of equations becomes: \\

\begin{equation}
\begin{aligned}
\frac{dn}{dt} &= \mu (1-p) - \beta sn - \mu n \\
\frac{ds}{dt} &= \beta sn - \alpha sa + \sigma qs - \mu s\\
\frac{dq}{dt} &= \alpha sa - \sigma qs - \gamma q - \mu q \\
\frac{da}{dt} &= \mu p + \gamma q - \mu a
\end{aligned}
\label{eq:system2}
\end{equation}

Using the fact that $n = 1-s-q-a$ and reduce the above system to the following:
\begin{equation}
\begin{aligned}
\frac{ds}{dt} &= \beta s(1-s-q-a) - \alpha sa + \sigma qs - \mu s\\
\frac{dq}{dt} &= \alpha sa - \sigma qs - q\gamma - q\mu \\
\frac{da}{dt} &= \mu p + \gamma q - \mu a
\end{aligned}
\label{eq:system3}
\end{equation}

The model system~(3) always admits a \emph{smoking-free equilibrium}
\(
E_0 = (0,0,p),
\)
at which the entire population consists of non-smokers. \\

In addition, system~(3) possesses an \emph{endemic equilibrium}
\(
E^* = (s^*, q^*, a^*),
\)
whose explicit form is more involved. By setting
\[
\frac{\mathrm{d}s}{\mathrm{d}t} = 0, \qquad
\frac{\mathrm{d}q}{\mathrm{d}t} = 0, \qquad
\frac{\mathrm{d}a}{\mathrm{d}t} = 0,
\]
where the components of $E^*$ can be expressed as
\[
a^* = p + \frac{\gamma}{\mu} q^*, 
\qquad
q^* = \frac{\alpha s^* a^*}{\sigma s^* + \gamma + \mu},
\]
together with the constraint
\[
\beta (1 - s^* - q^* - a^*) - \alpha a^* + \sigma q^* - \mu = 0.
\]

Substituting the expression for $q^*$ into that for $a^*$ yields
\[
a^* = p + \frac{\alpha \gamma s^* a^*}{\mu(\sigma s^* + \gamma + \mu)}.
\]
Solving for $a^*$ in terms of $s^*$ gives
\[
a^* = \frac{p \mu (\sigma s^* + \gamma + \mu)}
           {s^* (\mu \sigma - \gamma \alpha) + \mu (\gamma + \mu)}.
\]
Using this result in the expression for $q^*$, we obtain after simplification
\[
q^* = \frac{\alpha p \mu s^*}
           {s^* (\mu \sigma - \gamma \alpha) + \mu (\gamma + \mu)}.
\]

Substituting these expressions and simplifying, we obtain the following quadratic equation for $s^*$:
\[
\hat{A} (s^*)^2 + \hat{B} s^* + \hat{C} = 0,
\]
where
\begin{align*}
\hat{A} &= \beta (\gamma \alpha - \sigma \mu),\\
\hat{B} &= \mu (\beta \sigma + \gamma \alpha) - \mu^2 (\sigma + \beta)
     - \mu \beta p (\alpha + \sigma) - \beta \gamma (\mu + \alpha),\\
\hat{C} &= \mu (\gamma + \mu)(\mu + \alpha p)(\mathcal{R} - 1).
\end{align*}

For $\hat{A} > 0$, we require
\[
\beta (\gamma \alpha - \sigma \mu) > 0,
\]
which implies
\[
\gamma \alpha - \sigma \mu > 0,
\qquad \text{or equivalently} \qquad
\frac{\gamma \alpha}{\sigma \mu} > 1.
\]
The signs of $b$ and $c$ may vary depending on the parameter values. In particular, the sign of $c$ is determined by that of $(\mathcal{R} - 1)$, where
\[
\mathcal{R} = \frac{\beta (1 - p)}{\mu + \alpha p}.
\]

\subsection{The Basic Reproduction Number}

The quantity $\mathcal{R}$ plays a role that is analogous to the basic reproduction number in epidemic models \cite{Diekmann1990}. For the model system (1) the basic reproduction number is: 
\[ \mathcal{R} = \frac{\beta (1-p)}{\mu + \alpha p}. \]
The value of $\mathcal{R}$ depends upon many parameters. To have a clear notion about the effect of various parameters on the value of R, we examine the sensitivity of each parameter on it. The normalised sensitivity indices are obtained as follows: 
\[ \Gamma_{\beta}^{\mathcal{R}} = 1, \Gamma_{p}^{\mathcal{R}} = \left|-\frac{(\mu + \alpha)}{(\mu /p+ \alpha)(1-p)} \right|, \Gamma_{\alpha}^{\mathcal{R}} = \left|-\frac{p\alpha}{\beta (\mu + \alpha p) (1-p)}\right|, \Gamma_{\mu}^{\mathcal{R}} = \left|-\frac{\mu}{\beta (\mu + \alpha p) (1-p)}\right|\]

Note that $\mathcal{R}$ is linearly sensitive to $\beta$, highlighting the dominant role of social influence or contact rate in driving the spread of smoking behavior. In contrast, increasing $p$ (initial awareness), $\alpha$ (quitting rate), or $\mu$ (influx rate) reduces $\mathcal{R}$, as indicated by their negative elasticities. The magnitudes of $\Gamma_{p}^{\mathcal{R}}$, $\Gamma_{\alpha}^{\mathcal{R}}$, and $\Gamma_{\mu}^{\mathcal{R}}$ quantify the relative strength of these suppressive effects. 

\subsection{Stability Analysis}
The Jacobian matrix at the Smoking-Free Equilibrium (SFE) is given by: \\

\[
J_{E_{0}} = 
\begin{bmatrix}
  \beta (1-p) - \alpha p - \mu & 0 & 0 \\
  \alpha p  & -\gamma - \mu & 0 \\
  0 & \gamma & -\mu \\
\end{bmatrix}
\]

The eigenvalues for this are trivial to calculate. Two of them are always negative: $-(\gamma+\mu),$ $-\mu$. The other one is negative for  $ \mathcal{R} > 1.$ Note that if $ \mathcal{R} < 1,$ all eigenvalues of $J_{E_{0}}$ are negative whereas one eigenvalue becomes positive if $ \mathcal{R} > 1.$ Thus, the SFE is locally asymptotically stable if $ \mathcal{R} < 1$ and unstable (saddle point) if $ \mathcal{R} > 1.$ \\

The local stability of the endemic equilibrium $E^*$ is determined by the eigenvalues of the Jacobian matrix evaluated at $E^*$, given by \\

\[
J_{E^*} =
\begin{pmatrix}
-\beta s^* & -\beta s^* + \sigma s^* & -\alpha s^* - \beta s^* \\
\alpha a^* - \sigma q^* & -\sigma s^* - \gamma - \mu & \alpha s^* \\
0 & \gamma & -\mu
\end{pmatrix}.
\]

The associated characteristic equation of $J_{E^*}$ is given by
\[
\lambda^3 + A_1 \lambda^2 + A_2 \lambda + A_3 = 0,
\]
where
\[
\begin{aligned}
A_1 &= \beta s^* + \sigma s^* + \gamma + 2\mu, \\[4pt]
A_2 &= \beta s^* \mu + (\mu + \beta s^*)(\sigma s^* + \gamma + \mu)
      + s^*(\beta - \sigma)(\alpha a^* - \sigma q^*)
      - \gamma \alpha s^*, \\[4pt]
A_3 &= \alpha^2 \gamma a^* s^* + \alpha \beta \gamma a^* s^*
       + \alpha \beta \mu a^* s^* - \alpha \mu \sigma a^* s^* - \alpha \beta \gamma (s^*)^2 - \alpha \gamma \sigma q^* s^*
       - \beta \gamma \sigma q^* s^* - \beta \mu \sigma q^* s^* \\
    &\quad + \beta \mu (s^*)^2 \sigma + \mu q^* s^* \sigma^2
       + \beta \gamma \mu s^* + \beta \mu^2 s^* .
\end{aligned}
\]

By the Routh--Hurwitz criterion, all eigenvalues have negative real parts if and only if
\[
A_2 > 0, \qquad A_3 > 0, \qquad A_1 A_2 > A_3.
\]
Hence, the endemic equilibrium $E^*$ is locally asymptotically stable whenever these conditions are satisfied. 

The following theorems summarize the results concerning the existence and stability of the equilibria of the model system~(3).

\begin{theorem}
If $\mathcal{R}<1$, the smoking-free equilibrium $E_0$ exists and is locally asymptotically stable. 
\end{theorem}

\begin{theorem}
When $\mathcal{R}>1$, the smoking-free equilibrium $E_0$ becomes unstable and an endemic equilibrium $E^*$ exists.
The endemic equilibrium $E^*$ is locally asymptotically stable if and only if the following conditions hold:
\[
A_2 > 0, \qquad A_3 > 0, \qquad A_1 A_2 > A_3.
\]
\end{theorem}

\section{Numerical Simulations}

This section presents numerical simulations of model system~(3), performed using Python~3.8.5. The parameter values employed in the simulations are
\[
\beta = 0.7,\quad \mu = 0.12,\quad \alpha = 0.5,\quad \sigma = 0.01,\quad \gamma = 0.042,\quad p = 0.34.
\]
For these values, the quadratic equation determining $s^*$ has coefficients $0.01386$, $-0.0396576$, and $0.00334366$, and the basic reproduction number is $R = 1.59$. The corresponding equilibrium point is
\[
(s^*, q^*, a^*) = (0.0869563,\, 0.1001176,\, 0.3750411).
\]

Figure 2 illustrates the global stability of the endemic equilibrium in the $s-q-a$ space. Trajectories are initiated from a variety of different initial conditions, and all converge to the same equilibrium point. This consistent convergence provides numerical confirmation that the endemic equilibrium is globally stable within the domain of feasible population fractions. 
Further, the effect of key model parameters on the time evolution of fractions of different populations in the deterministic compartmental framework is shown in Figure 3. Specifically, we examine the effects of $\beta$, $\alpha$, and $\sigma$ on the long-term equilibrium fractions of smokers ($S$) and quitters ($Q$) in the deterministic compartmental model. The results show that increasing $\beta$ leads to a higher equilibrium number of smokers, while increasing $\alpha$ reduces the number of smokers. In contrast, increasing $\sigma$ decreases the equilibrium number of quitters but increases the number of smokers. These trends highlight the differential roles of social influence ($\beta$), cessation rate ($\alpha$), and relapse rate ($\sigma$) in shaping the steady-state behavior of the system.

\begin{figure}
\centering
\includegraphics[width=0.9\textwidth]{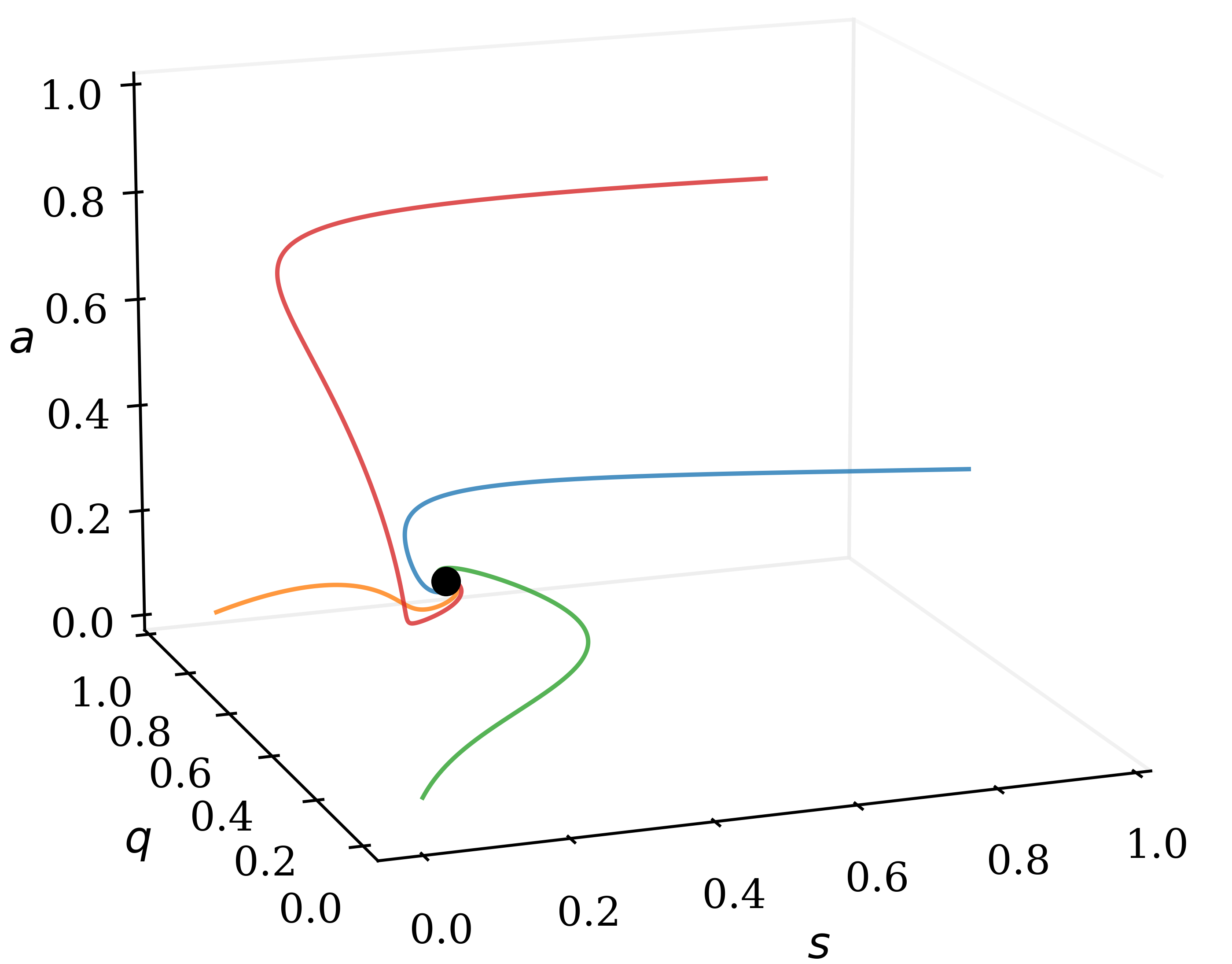}
\caption{Global stability of the endemic equilibrium in the $(s,q,a)$ space. Trajectories starting from diverse initial conditions all converge to the same equilibrium point, providing numerical evidence of the global stability of the system.}
\label{fig:global-stability}
\end{figure}

\begin{figure}
\centering
\includegraphics[width=\textwidth]{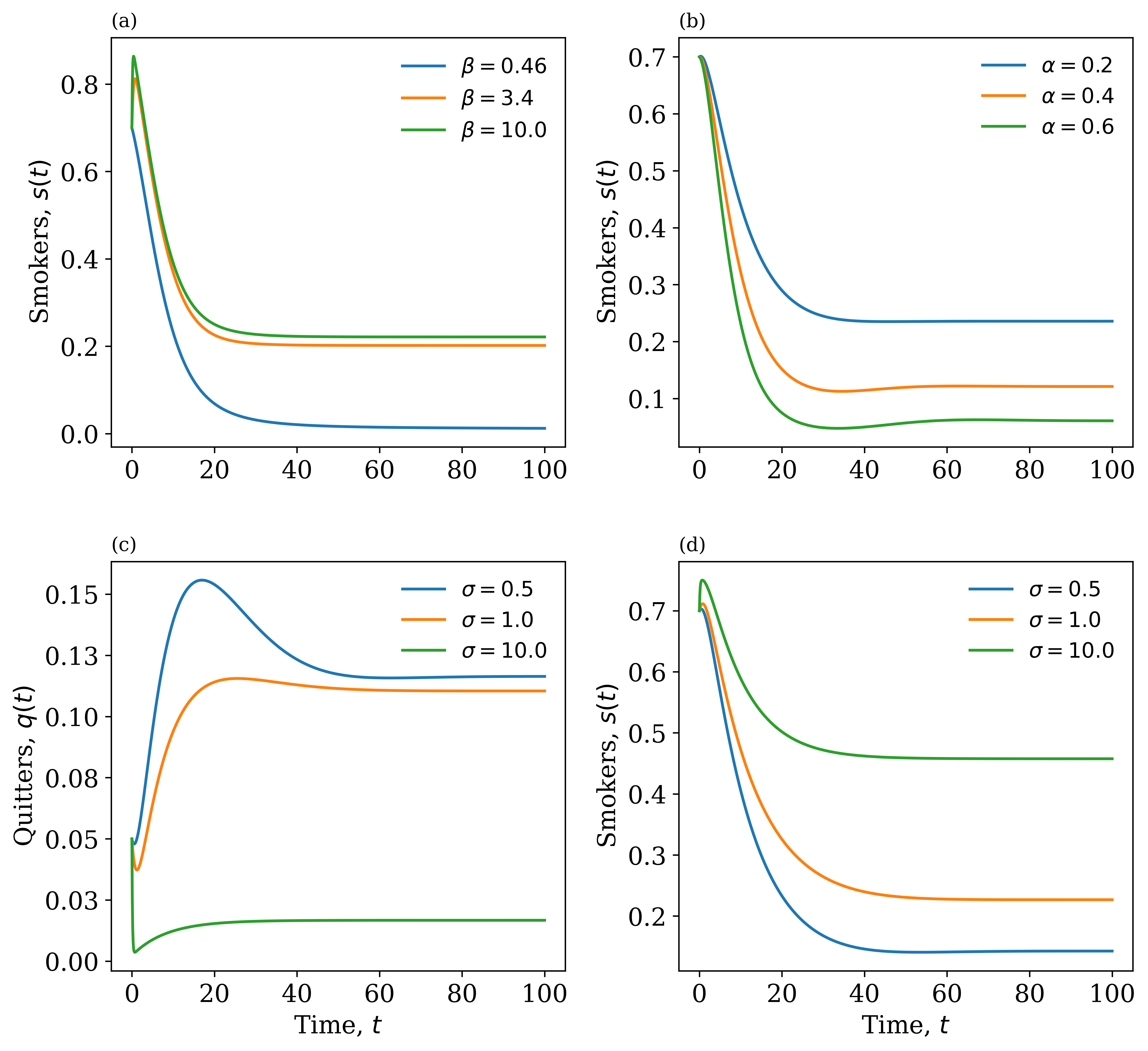}
\caption{Effects of model parameters on quitter and smoker dynamics in the ODE model.}
\label{fig:ODE-2x2}
\end{figure}

\section{Network-Based Modelling}

The compartmental model provides an accurate view of {\em long-term} dynamics, however capturing the spread of smoking behavior in a population where each individual interacts with very different neighborhoods requires a heterogeneous approach. To address this, we turn to network-based modeling. The role of peer influence in smoking uptake and cessation is well documented in empirical studies. Classic work by Christakis and Fowler (2008) analyzed a large social network and found that if a person quits smoking, the chances that their friends quit increase significantly, leading to clusters of quitting spreading through the network. Similarly, smoking initiation often occurs in friendship groups, where social norms and peer pressure play a decisive role. Adolescents with a higher proportion of smoking friends are far more likely to start smoking themselves, and vice versa for those with non-smoking friends-\cite{Christakis2008}. These phenomena imply that the network structure of social interactions can critically shape smoking prevalence over time. Dense clusters of smokers can sustain each other’s behavior, while if a cluster shifts towards non-smoking, that norm can reinforce itself. Hence, we introduce a network-based model that explicitly captures how peer influence shapes both smoking uptake and cessation. \\

%
\subsection{Degree block approximation}

Consider the network-based version of the model, where $n_k(t)$, $s_k(t)$, $q_k(t)$, and $a_k(t)$ the fractions of nodes of degree $k$ in the
corresponding compartments.
These variables satisfy the normalization condition
\begin{equation*}
n_k + s_k + q_k + a_k = 1 .
\end{equation*}
Nonlinear interaction terms are approximated by assuming that the state of a
neighbor of a degree-$k$ node is independent of $k$ and is determined only by
the degree distribution. In particular, the probability that a randomly chosen
neighbor is in compartment $x$ is given by
\begin{equation*}
\Theta_x = \frac{\sum_j j P(j) x_j}{\langle k \rangle}, 
\qquad x \in \{s,q,a\},
\end{equation*}
where $P(j)$ is the degree distribution and $\langle k \rangle$ is the mean
degree. With these definitions, the degree-block approximation of system
\begin{equation}
\begin{aligned}
\frac{dn_k}{dt} &=
\mu (1-p) - \beta \, k \, n_k \, \Theta_s - \mu n_k, \\[2mm]
\frac{ds_k}{dt} &=
\beta \, k \, n_k \, \Theta_s - \alpha \, k \, s_k \, \Theta_a + \sigma \, k \, q_k \, \Theta_s - \mu s_k, \\[1mm]
\frac{dq_k}{dt} &=
\alpha \, k \, s_k \, \Theta_a - \sigma \, k \, q_k \, \Theta_s - (\gamma + \mu) \, q_k, \\[1mm]
\frac{da_k}{dt} &=
\mu p + \gamma q_k - \mu a_k,
\end{aligned}
\end{equation}
In the degree-based formulation, the parameters $\alpha$, $\beta$, and $\sigma$
are interpreted as \emph{per-contact} transition rates. A node of degree $k$
interacts with $k$ neighbors, so that the corresponding transition terms are
proportional to $k\Theta_x$, where $\Theta_x$ denotes the probability that a
randomly chosen neighbor is in compartment $x$. The well–mixed model is recovered
in the homogeneous limit by replacing $k$ with the mean degree $\langle k\rangle$.
In this case, the effective transition rates become
$\alpha\langle k\rangle$, $\beta\langle k\rangle$, and $\sigma\langle k\rangle$,
and the degree-block system reduces to the original mean–field model. Thus, the
network model generalizes the well–mixed system by explicitly accounting for
degree heterogeneity while preserving the same underlying transition mechanisms.

\subsection{Numerical simulation on Theoretical Networks}
To investigate how peer network structure shapes smoking behavior, we employed custom code to simulate these dynamics using the Gillespie Stochastic Simulation Algorithm (SSA)~\cite{Gillespie1977}. Simulations were conducted across three canonical network types—Erdős-Rényi (random), Albert-Barabási (scale-free), and Watts-Strogatz (small-world)—to capture the effects of different network topologies on behavioral propagation. Each simulation was repeated 100 times. We plotted the mean trajectory of smoking adoption over time. Shaded bands representing the standard deviation from the mean were included to illustrate the variability in behavior propagation across different network structures. In these simulations, the parameter values taken for the simulations are: $\beta = 0.7$, $\alpha = 0.5$, $\sigma = 0.01$, $\gamma = 0.042$. The average degree of the network is $\langle k \rangle = 8,$ and the initial conditions used in all network simulations are 
\(
(N(0), S(0), Q(0), A(0)) = (0.8, 0.005, 0, 0.195).
\)

To provide context for our choice of networks, we briefly describe the technical properties of each network model used in the simulations:

\begin{itemize}
\setlength{\itemsep}{1em}
 \item[1.] \textbf{Erdős-Rényi Network}: This network is generated by connecting nodes randomly, with each pair of nodes having an equal probability \( p \) of being connected~\cite{ErdosRenyi1960}. 
 \item[2.] \textbf{Albert-Barabási Network (Scale-Free Network)}: This network exhibits a power-law degree distribution, where a small number of nodes (hubs) have a disproportionately high number of connections~\cite{BarabasiAlbert1999}. 
 \item[3.] \textbf{Watts-Strogatz Network (Small-World Network)}: The Watts-Strogatz network is generated by rewiring a regular lattice with a probability \( p \), which introduces random shortcuts while preserving local clustering~\cite{WattsStrogatz1998}. 
\end{itemize}

\begin{figure}
\centering
\includegraphics[width=\textwidth]{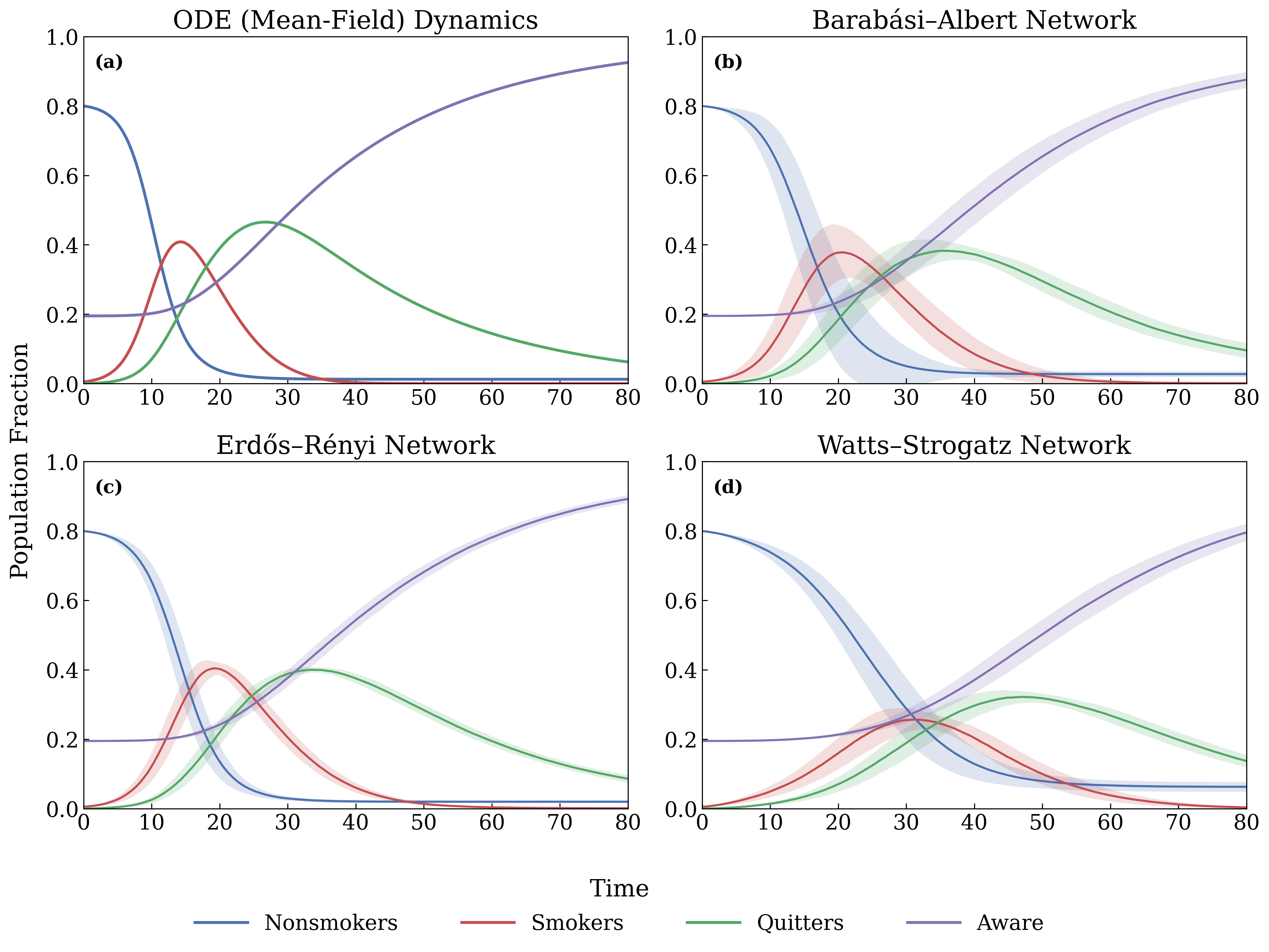}
\caption{Comparison of the temporal dynamics under the deterministic mean-field model and network-based simulations. 
(a) Deterministic compartmental (ODE) model. 
(b) Barab\'asi--Albert Network. 
(c) Erdős-Rényi Network. 
(d) Watts-Strogatz Network. 
While quantitative differences are observed in the timing of peak prevalence and in the shape of the smoking trajectory across network topologies, the qualitative behavior and long-term dynamics remain consistent with the mean-field predictions.}
\label{fig:hom-vs-het}
\end{figure}

We implemented our model on these three networks, and found that the overall dynamical behavior remains consistent across different network topologies and closely resembles that of the ODE-based compartmental model, as shown in Figure \ref{fig:hom-vs-het}. We observed quantitative differences, such as variations in the timing of the epidemic peak and differences in the shape of the smoking prevalence curve (e.g., sharper versus flatter trajectories), however the qualitative behavior of the system remains largely unchanged. In particular, the long-term dynamics and overall trends agree with those predicted by the deterministic mean-field framework.

\begin{figure}
\centering
\includegraphics[width=\textwidth]{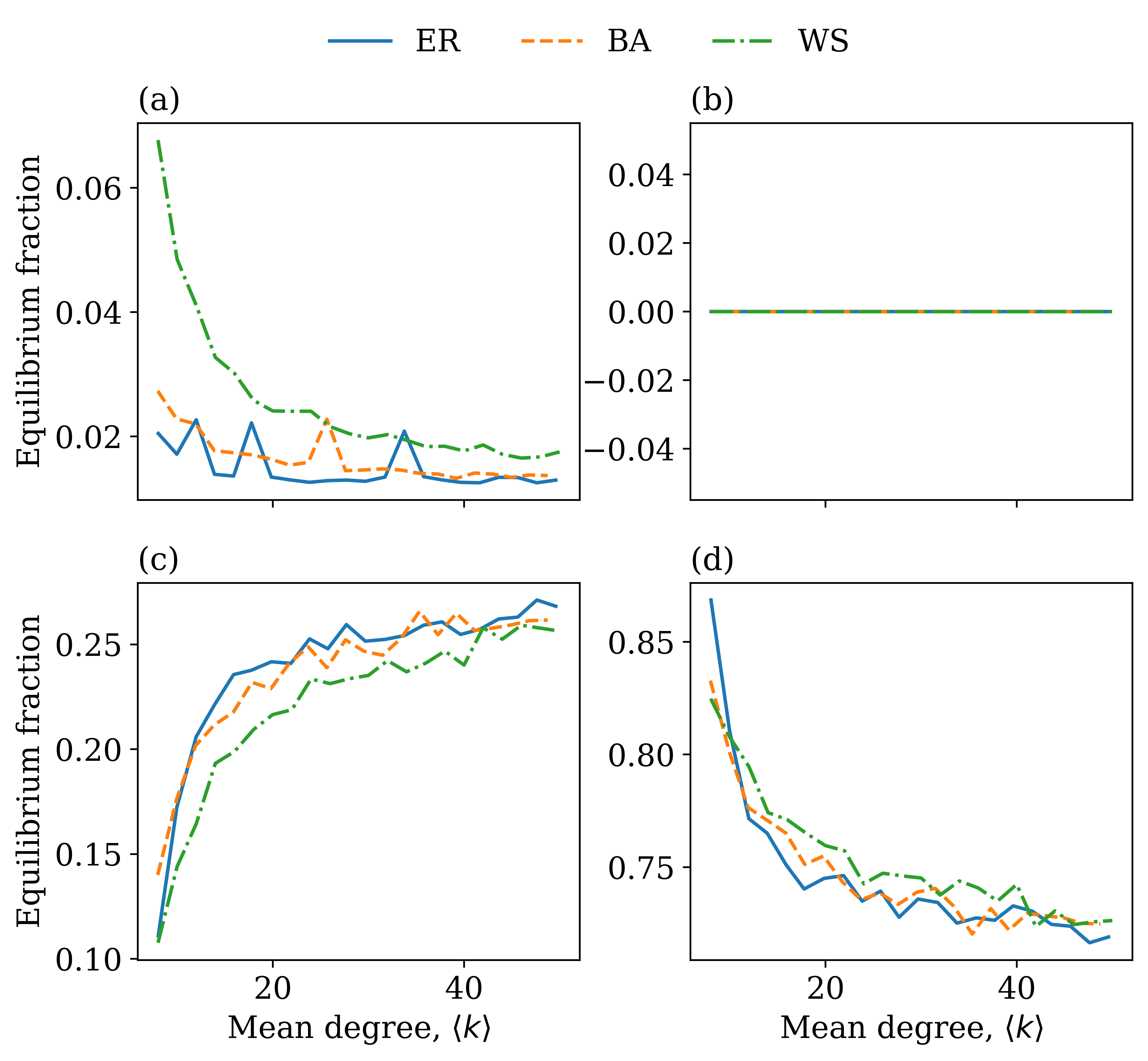}
\caption{Equilibrium population fractions as a function of mean degree $\langle k\rangle$ for Erdős–Rényi (ER), Barabási–Albert (BA), and Watts–Strogatz (WS) networks. Panels show (a) nonsmokers, (b) smokers, (c) quitters, and (d) aware individuals.}
\label{fig:equilibrium-fractions}
\end{figure}

To further quantify the impact of network connectivity on the long-term system behavior, we analyzed how the equilibrium fractions of the non-smoker ($N$), smoker ($S$), aware ($A$), and quitter ($Q$) compartments depend on the average network degree. We plotted the equilibrium population levels as functions of the mean degree for the different network topologies considered. Since the population is closed, the equilibrium fraction of smokers satisfies $S(t) \to 0$ as $t \to \infty$. However, the remaining compartments exhibit a clear dependence on the average degree. The dependence of the equilibrium fractions on the average network degree is illustrated in Figure~\ref{fig:equilibrium-fractions}. Note that the equilibrium fraction of non-smokers decreases as the mean degree increases, whereas the fraction of quitters increases with increasing connectivity. In contrast, the equilibrium level of awareness decreases as the average degree grows. These trends are consistent across all theoretical network structures examined in this study.

\subsection{Numerical simulation on Empirical Networks}
We will analyze three empirical networks: Facebook SNAP Dataset, Hamsterster Dataset and Rovira University Dataset.
The table below outlines key network parameters such as the number of nodes and edges, average degree, density, and clustering coefficient. The {\em average degree} provides an indication of the typical connectivity of nodes within each network. {\em Density} reflects how close the network is to being fully connected, while the {\em clustering coefficient} measures the tendency of nodes to form tightly knit clusters. 

\begin{table}
\centering
\caption{Statistical Summary of Empirical Networks}
\renewcommand{\arraystretch}{1.25}
\begin{tabular}{|c|c|c|c|c|c|}
\hline
\textbf{Network} & \textbf{Nodes} & \textbf{Edges} & \textbf{Average Degree} & \textbf{Density} & \textbf{Clustering Coefficient} \\
\hline
Facebook & 4039 & 88234 & 43.69 & 0.0108 & 0.6055 \\
Hamsterster & 2426 & 16631 & 13.71 & 0.0057 & 0.5376 \\
Rovira Uni & 1133 & 5451 & 9.62 & 0.0085 & 0.2202 \\
\hline
\end{tabular}
\end{table}

\begin{figure}[htbp]
\centering

\begin{overpic}[width=0.48\textwidth]{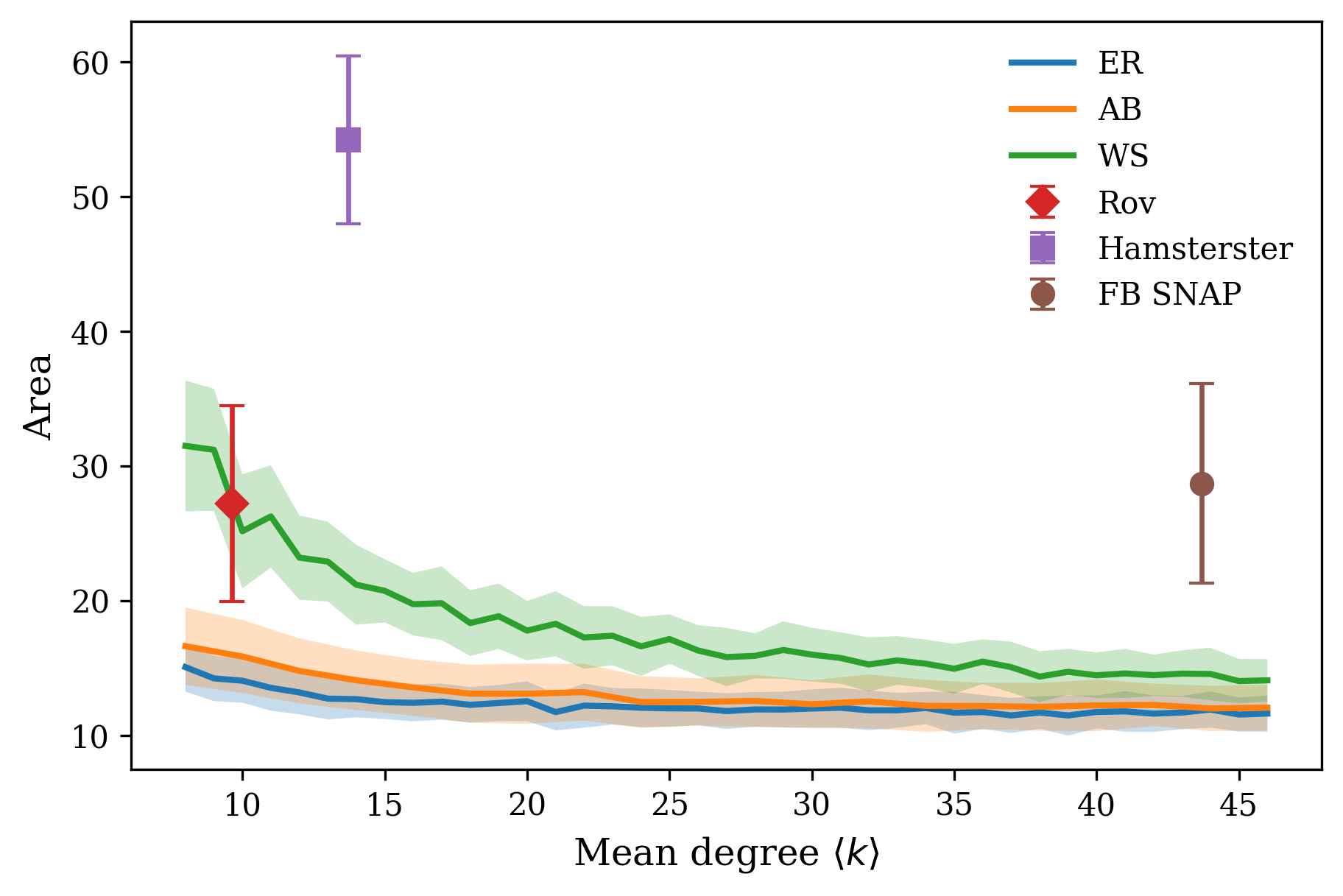}
\put(2,67){\textbf{(a)}}
\end{overpic}
\hfill
\begin{overpic}[width=0.48\textwidth]{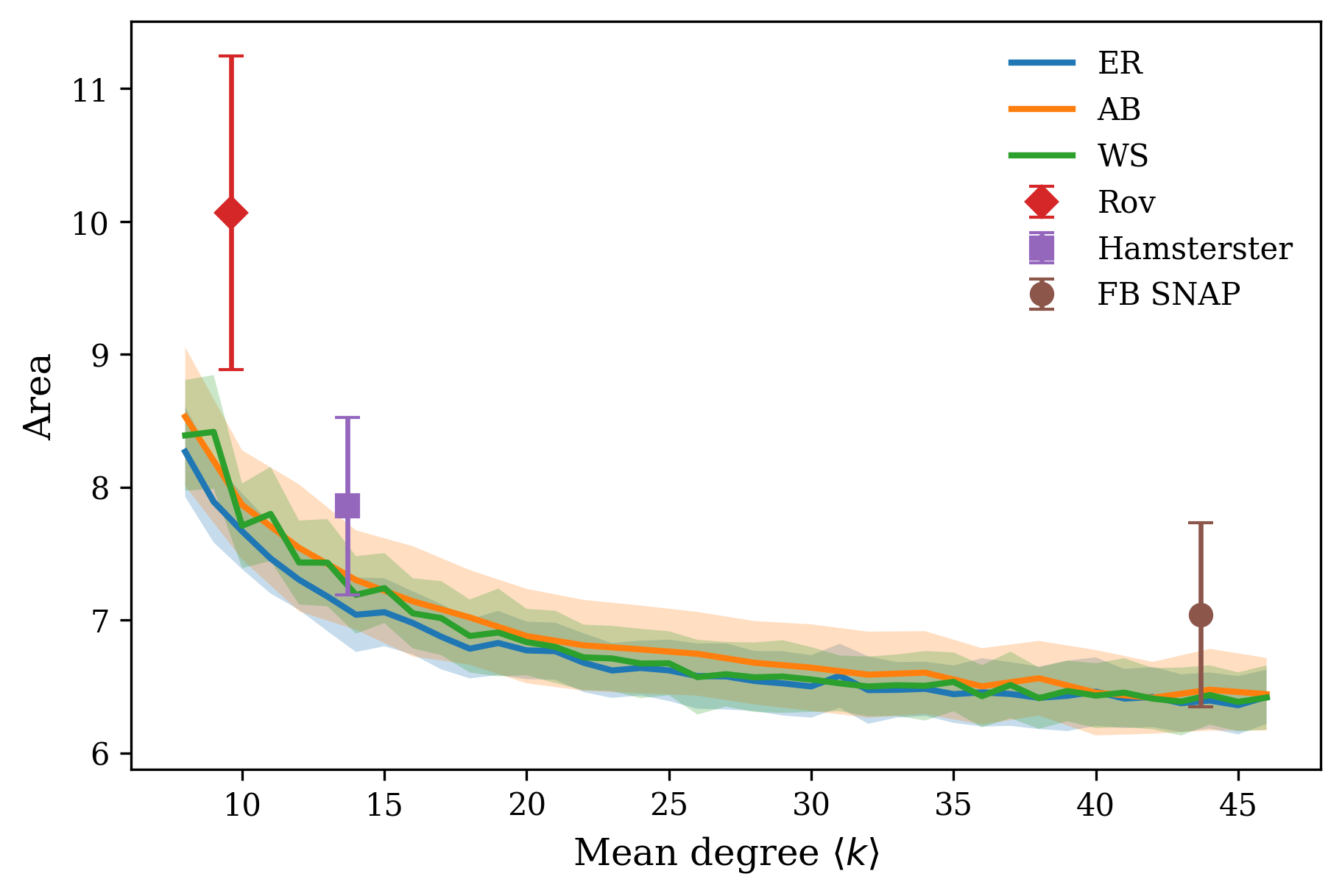}
\put(2,67){\textbf{(b)}}
\end{overpic}

\vspace{0.4cm}

\begin{overpic}[width=0.48\textwidth]{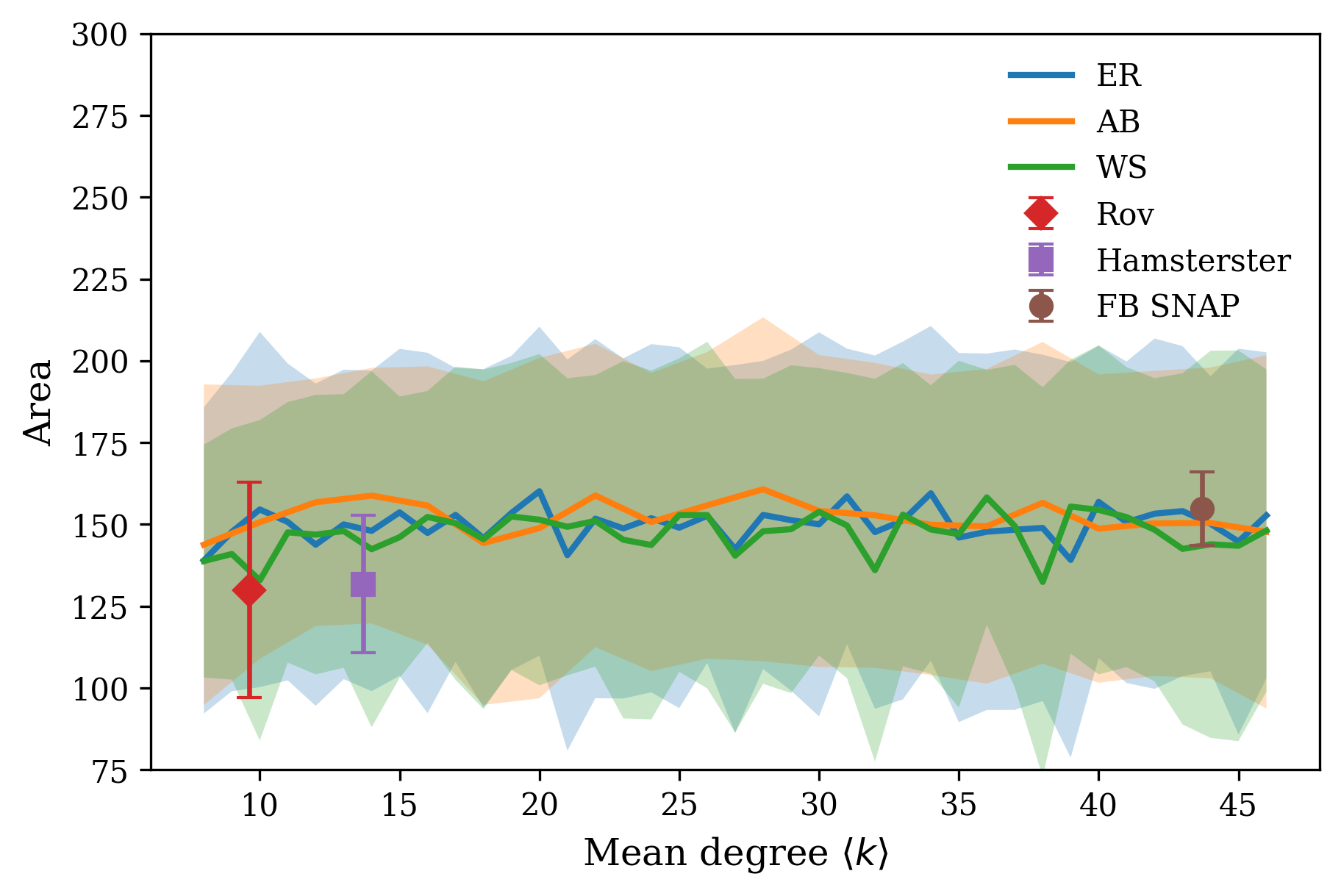}
\put(2,67){\textbf{(c)}}
\end{overpic}
\hfill
\begin{overpic}[width=0.48\textwidth]{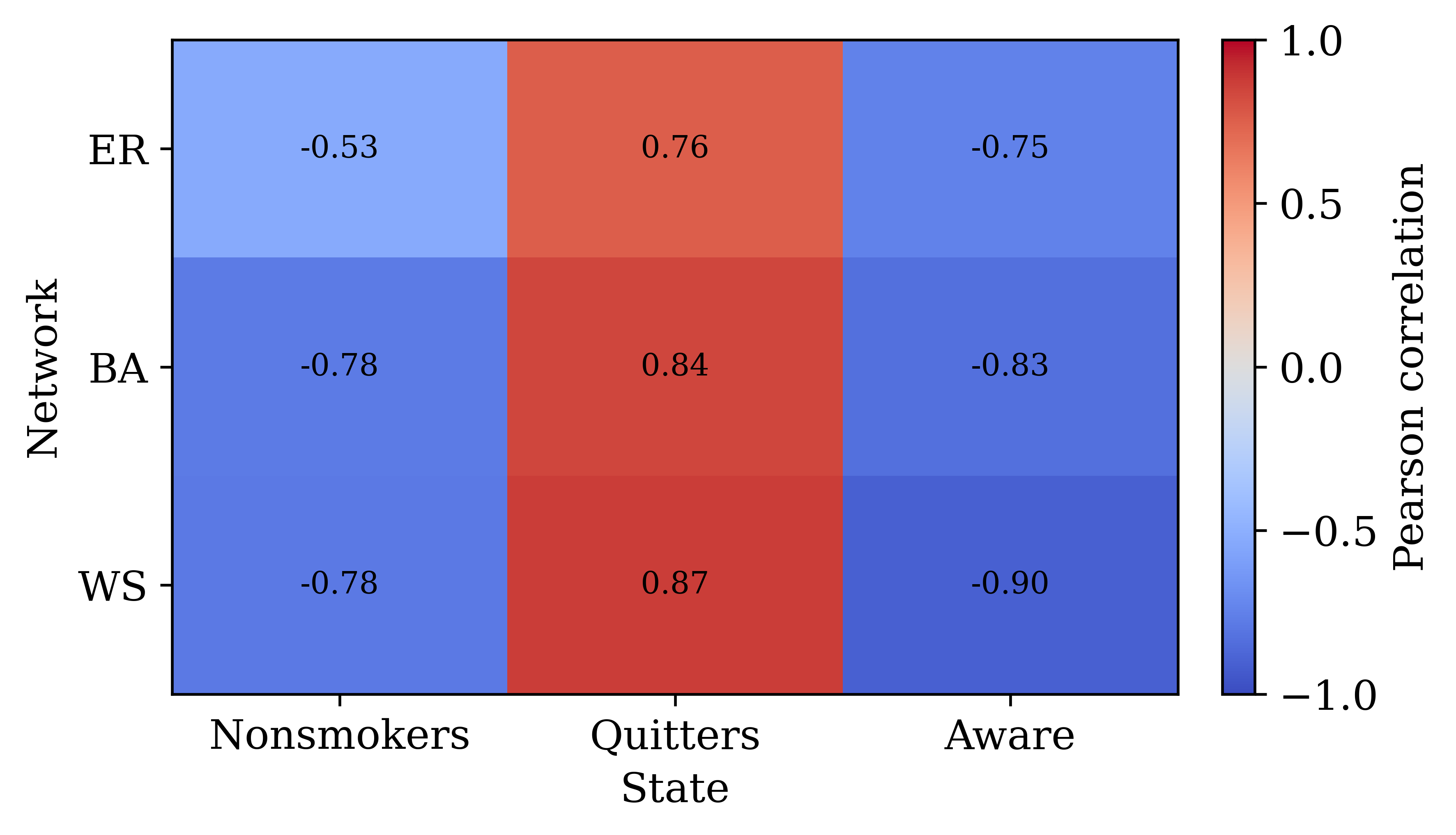}
\put(2,67){\textbf{(d)}}
\end{overpic}

\caption{Dependence of cumulative population on mean network degree. 
(a) Area under nonsmokers curve. 
(b) Area under smoker curve. 
(c) Area under aware curve. 
(d) Pearson correlation between $\langle k\rangle$ and equilibrium fractions across network topologies.}
\label{fig:combined}
\end{figure}

To quantify the cumulative dynamics of the system, we computed the area under the temporal curves of the non-smoker ($N$), smoker ($S$), and aware ($A$) populations for each simulation. This measure captures the total number of individuals in each compartment over time, complementing the equilibrium analysis. The calculations were performed across three theoretical network models (Erd\H{o}s--R\'enyi, Barab\'asi--Albert, and Watts--Strogatz) as well as three empirical social networks, and are shown in Figure~\ref{fig:combined}. Note that, as the average degree $\langle k \rangle$ increases, the cumulative fractions of non-smokers and smokers decrease, whereas the cumulative number of aware individuals remains constant. This demonstrates that, while network degree strongly influences smoking and quitting dynamics, the total number of individuals exposed to awareness over the course of the dynamics is largely independent of connectivity. The independence of the cumulative awareness level from the network degree can be explained by the underlying transition mechanism in the model. In our framework, individuals do not become aware directly through contact with aware individuals. Instead, awareness is acquired only after an individual quits smoking. That is, quitting precedes awareness, and aware individuals subsequently contribute to awareness dissemination. Because the transition into the aware compartment is not driven by direct pairwise interactions but rather by internal state progression following cessation, the cumulative number of aware individuals does not explicitly depend on network connectivity. In contrast, smoking adoption and quitting are contact-driven processes and therefore remain sensitive to the average degree of the network.
Hence, these findings emphasize the distinction between cumulative outcomes and equilibrium fractions.

\section{Discussion}

In this study, we analyze the behavioral mechanisms underlying smoking and smoking cessation, focusing on the social and awareness-driven factors that shape individual decisions. Through the lens of a proposed compartmental contagion model, we examine how local peer influence within social networks and broader awareness effects interact to drive behavioral transitions between smokers, quitters, and aware individuals. By combining analytical investigation with network-based modeling, we identify the key conditions under which awareness can mitigate or suppress the persistence of smoking. A central contribution of this work lies in the development of a modeling framework that integrates social structure and behavioral feedback, enabling a systematic analysis of smoking dynamics at both the population and network levels.

We first propose and analyze a deterministic compartmental model consisting of four classes: non-smokers(N), smokers ($S$), quitters ($Q$), and aware individuals ($A$). The stability results of this model indicate that the long-term dynamics of smoking behavior depend critically on the threshold parameter $\mathcal{R}$. When $\mathcal{R}<1$, the Smoking-Free Equilibrium is locally asymptotically stable, implying that smoking cannot persist in the population and any small introduction of smokers will eventually die out. In contrast, when $\mathcal{R}>1$, the smoking-free state becomes unstable and the system admits a stable endemic equilibrium $E^*$, corresponding to the persistence of smoking in the population at a constant level. This highlights the importance of reducing key transmission and initiation parameters (such as $\beta$ and $p$) or increasing cessation-related parameters (such as $\alpha$ and $\gamma$) in order to bring $\mathcal{R}$ below unity and eliminate smoking from the population. In addition, relapse (parameter $\sigma$) plays a decisive role in sustaining smoking: higher relapse increases the long-run smoker fraction and sharply reduces the quitter fraction. This highlights an important practical implication: interventions that promote quitting are likely to be substantially more effective when paired with relapse-prevention supports (e.g., sustained cessation assistance), because relapse provides a feedback pathway that can repopulate the smoker compartment even after awareness-driven declines. 

To further examine the effects of heterogeneity, contact patterns, and network structure on the coupled dynamics of awareness spread and smoking adoption, we implemented our model on three distinct network topologies: a random network, a Barab\'asi--Albert scale-free network, and a small-world network. These network structures exhibit significantly different connectivity patterns and degree distributions. Our simulation results are consistent with those obtained from the deterministic compartmental model. However, to more thoroughly investigate the role of network degree, we analyzed how variations in average degree influence the system dynamics. The network-based analysis indicates that the prevalence of awareness is largely independent of the network degree, whereas the number of smokers and quitters depends strongly on the average degree of the network. These network-based results provide important policy implications. Since awareness prevalence appears largely independent of network degree, mass awareness campaigns can be effective across diverse social structures. However, because smoking and quitting dynamics depend strongly on network connectivity, highly connected populations are more susceptible to smoking propagation. Therefore, public health strategies should combine broad awareness campaigns with targeted interventions in high-contact environments and among highly connected individuals. Incorporating network structure into policy design may substantially enhance the effectiveness and efficiency of smoking control programs.

 \bibliography{bibpp1}
 \bibliographystyle{elsarticle-num}

\end{document}